\documentstyle[PASJadd]{PASJ95}
\begin{document}

\title{Intermittent Star-Formation Activities of Dwarf Irregular
Galaxies} 
\author{Hiroyuki {\sc Hirashita}\thanks{%
Research Fellow of the Japan Society for the Promotion of Science.}
\\[12pt]
{\it Department of Astronomy, Faculty of Science, Kyoto University,
Sakyo-ku, Kyoto 606-8502}\\
{\it E-mail(HH): hirasita@kusastro.kyoto-u.ac.jp}}

\abst{On the basis of the propagating star-formation model,
we investigated the star-formation activities of dwarf irregular
galaxies (dIrrs) by considering two processes: the heating
from stellar feedback and the cooling of the heated
gas. After examining the timescales of the two processes, we propose
that continuous global star-formation activity is difficult
in dIrrs, 
since their small sizes make the feedback efficient and
their small metallicities prevent the cooling from becoming
effective. Thus, the intermittent nature of the star-formation
activities of dIrrs, which is due to the small
metallicity as well as the small size, is predicted.
We emphasize that the size of a galaxy is an important factor
concerning star-formation activity.
The intermittence of the star-formation activity is also supported
by the observed scatter of the {\it UBV} colors and H$\alpha$
equivalent widths of the dIrr sample. However, we note that
efficient interstellar mixing may make the
cooling time much shorter.
}

\kword{galaxies: dwarf --- galaxies: evolution --- galaxies: irregular --- H II regions --- ISM: bubbles}

\maketitle
\thispagestyle{headings}

\section{Introduction}

Dwarf irregular galaxies (dIrrs) are small galaxies characterized by
their ``irregular morphologies'' and their star-formation
activities. Their absence of spiral structures makes it
difficult for us to understand their star-formation activities
based on the density wave theory (e.g., Spitzer 1978). Thus,
there is a motivation different from the density-wave theory to
investigate the activities in dIrrs.

One of the models for star-formation activity is
``self-propagating'' star formation. Historically, this model
was developed to explain the structures of spiral galaxies
(Mueller, Arnett 1976; Gerola, Seiden 1978). It
is examined by computer simulations under the assumption that
a local star-formation activity propagates to its neighboring
regions stochastically. The triggering mechanism of neighboring
star formation is generally thought to be compression by
shocks from stellar mass loss (Elmegreen, Elmegreen 1978).
In the model of self-propagating star formation,
spiral patterns emerge due to an interplay between the
propagation of star formation and the differential rotation
of the galactic disk. The spiral pattern is also understood to
be a dissipative structure (Nozakura, Ikeuchi 1988).

Another motivation for self-propagating star formation is
application to the Large Magellanic Cloud
(Feitzinger et al.\ 1981; Kamaya 1998).
Dopita et al.\ (1985) showed observationally that
the star formation in the Shapley Constellation III
region of the Cloud propagates at a velocity of
36 km s$^{-1}$. A general application to dwarf galaxies was
described in Gerola et al.\ (1980). The nature of dwarf galaxies
is the main topic of this paper. We note that the picture of the
propagating star formation is simply applied to the
estimate of the propagating timescale of the feedback from the
star-formation activities (subsection 2.1). In this paper,
``stellar feedback''
means the heating effect by stars (e.g., supernovae, stellar
wind, ultraviolet radiations from OB stars; thermal conduction is
also considered in subsection 2.1). Since massive
stars with short lifetimes contribute to the effect most
significantly, we can consider that the feedback becomes
effective soon after the emergence of star-formation activity.

Some observational studies of star-formation activities in
dIrrs indicate that cool gas (e.g., Sait\={o} et al.\ 1992)
induces star-formation activity. Sait\={o} et al.\ (1992)
suggested that the infalling gas into a dwarf galaxy
induces star-formation activity through shock compression.
In this case, the radiative-cooling time of the heated gas is
of physical importance, since cooling is necessary for the
gas heated by supernovae to fall back into the galaxy
(Shapiro, Field 1976; Habe, Ikeuchi 1980). We
should note that the infalling gas may be primordial and
that the galaxy is still collapsing (Shostak, Skillman 1989).
However, we consider here that the gas originating from
the dIrr is cooled and induces the next star-formation activity.

In this paper, we examine the stellar feedback and
the cooling in dIrrs, both of which
are shown  above to be important (see also Cox, Smith 1974),
in the context of propagating star formation.
This paper is organized
as follows. In the next section (section 2) 
we consider the two processes and the intermittent nature of the
star-formation activities of dIrrs are explained. The final section
(section 3) is devoted to a summary and implications.

\section{Processes in Global Star-Formation Activities}

\subsection{Feedback from Star Formation}

There is observational evidence that a local star-formation
activity induces its neighboring star formations
(e.g., Dopita et al.\ 1985). The speed of the
propagation is $v_{\rm prop}\simeq 36$ km s$^{-1}$
(Dopita et al.\ 1985). The propagation is generally thought to
occur through supernova shock compression, which induces the next
star formation (for the physical processes, see e.g.,
Elmegreen, Elmegreen 1978).
Once star formation occurs in a region, that region is 
heated by supernovae, stellar winds, or ultraviolet radiations,
all of which suppress successive star formation in that
region. Considering that massive stars
with short lifetimes ($\ltsim 10^7$ yr) contribute to the
feedback most significantly, it is possible to consider
$v_{\rm prop}$ as being the propagation speed of stellar feedback
[this is justified by a comparison with equation (\ref{cross})].

We here adopt $v_{\rm prop}\simeq 36$ km s$^{-1}$ to empirically
estimate the propagation timescale of the stellar-feedback
effect. We assume that the star-formation activity (and the effect
of the feedback effect at almost the same time) propagates
at a speed of $v_{\rm prop}$. In fact, all shells of
superbubbles, which are responsible for the propagation, are not
expected to have the same speed. Thus, knowing the distribution of
the speed is necessary for a detailed study. However, we here
use the value 36 km s$^{-1}$ as a mean value of the expansion
speed. Indeed, the speed of the shell from an OB association is
typically on the order of 
10--100 km s$^{-1}$ (e.g., Weaver et al.\ 1977) and the
value 36 km s$^{-1}$ seems a reasonable choice.
Since $v_{\rm prop}$ is determined mainly by the expansion
speed of the superbubbles within a dwarf galaxy, the environment
does not seem to influence the value of $v_{\rm prop}$
(see e.g., Hirashita et al.\ 1997 for the environmental effects
on dwarf galaxies). In any case, we assume that
$v_{\rm prop}\simeq 36$ km s$^{-1}$ as a representative value.

The time evolution of hot gas
is also important, which has been extensively investigated in
Ferri\`{e}re (1998). We note that the treatment of an isolated
star-forming region in Ferri\`{e}re (1998) is different from ours,
since our assumption is that star formation propagates to its
neighbors and that star-forming regions correlates spatially.
Ferri\`{e}re concluded that the hot-gas filling factor of the
Galaxy is significantly less than 1. In this case, the feedback
from the star formation may not effective. Thus, we should mention
that the star-formation rate per unit volume in spatially
correlated star-forming regions should be
higher than that in the Galaxy for the effective feedback.
The physical treatment of the feedback is described in Spaans
and Norman (1997) and Scalo et al.\ (1998).

If the effect of stellar feedback propagates in the form
of a ``feedback wave'' with constant speed $v_{\rm prop}$,
the feedback affects the whole system in the
following crossing timescale ($t_{\rm cross}$) as
\begin{eqnarray}
t_{\rm cross} & \equiv & \frac{R}{v_{\rm prop}}\nonumber \\
& \simeq &
3\times 10^7\left(\frac{R}{1~{\rm kpc}}\right)\left(
\frac{v_{\rm prop}}{36~{\rm km~s}^{-1}}\right)^{-1}{\rm yr},
\label{cross}
\end{eqnarray}
where $R$ is the size of the system, which is estimated for
typical dIrrs.

Another empirical derivation of the crossing time is possible
based on the typical size of a giant H {\sc ii} region.
Star-forming dIrrs generally have
giant H {\sc ii} regions whose typical size is
$r_{\rm HII}\sim 100$ pc
(van den Bergh 1981; Hodge 1983; see also
Tomita et al.\ 1998). If an H {\sc ii} region propagates to
its neighboring region in its lifetime ($t_{\rm HII}\sim 10^7$ yr),
the crossing time of the propagating H {\sc ii} region is
\begin{eqnarray}
t_{\rm cross,\, HII} & = & \left(\frac{R}{r_{\rm HII}}\right)
t_{\rm HII}\nonumber \\
& \simeq & 10^8\left(\frac{R}{1~{\rm kpc}}\right)\left(
\frac{r_{\rm HII}}{100~{\rm pc}}\right)^{-1}\nonumber \\
& & \times\left(\frac{t_{\rm HII}}{10^7~{\rm yr}}\right)~{\rm yr}.
\label{cross2}
\end{eqnarray}
Comparing equations (\ref{cross}) and (\ref{cross2}), we
conclude that the crossing time of the star-formation activity
(feedback from star formation) is $10^7$--$10^8$
yr.

Once the hot gas is supplied by massive stars,
cold clouds suffers thermal conduction
(e.g., Draine, Giuliani 1984). According to
Cowie and McKee (1977), the timescale of evaporation due
to thermal conduction to cold clouds ($t_{\rm evap}$) is
estimated to be
\begin{eqnarray}
t_{\rm evap} & \simeq & 10^7\left(
\frac{n_{\rm cold}}{100\;{\rm cm}^{-3}}\right)\left(
\frac{R_{\rm cloud}}{1\;{\rm pc}}\right)^2\nonumber \\
& & \times\left(\frac{T_{\rm hot}}{10^6\;{\rm K}}\right)^{-5/2}
\left(\frac{\ln\Lambda}{30}\right)\;{\rm yr},\label{cond}
\end{eqnarray}
where $n_{\rm cold}$ and $R_{\rm cloud}$ are the number density
and the size of a cloud, respectively, $T_{\rm hot}$
is the temperature of the hot gas originating from the
stellar feedback, and $\ln\Lambda$ is the Coulomb logarithm,
which is a function of the electron number density and the electron
temperature of the hot gas (Spitzer 1956). Thus, the thermal
conduction is effective on the same timescale as the
crossing timescale. This means that the thermal conduction
can prevent star formation from cold clouds
(see also Hirashita 1999).

\subsection{Radiative Cooling and Dynamical Collapse}

The gas which lies near to a region with star-formation
activity tends to be heated by stellar feedback. The typical
temperature and number density of the heated gas are
$T\sim 10^6$ K and
$n\sim 10^{-3}$ cm$^{-3}$, respectively (McKee, Ostriker 1977).
Because the hot gas is tenuous, it expands in the dark-matter
halo (Tomisaka, Bregman 1993). To contribute to a star-formation
activity, the hot gas should cool and collapse.
Thus, the next star-formation emerges after the time interval
$t_{\rm int}$, estimated by
\begin{eqnarray}
t_{\rm int}={\rm max}(t_{\rm cool},~t_{\rm dyn}),\label{interval}
\end{eqnarray}
 where $t_{\rm cool}$ and
 $t_{\rm dyn}$ are the cooling timescale and the dynamical timescale
of the hot gas. Since the dynamics of the hot gas is
determined by the dark-matter potential, the dynamical timescale
(Binney, Tremaine 1987) is estimated to be
\begin{eqnarray}
t_{\rm dyn} & \simeq & \frac{R}{\sigma}\nonumber \\
& \simeq & 3\times 10^7\left(
\frac{R}{1~{\rm kpc}}\right)\left(\frac{\sigma}{30~{\rm km~s}^{-1}}
\right)^{-1}~{\rm yr},
\end{eqnarray}
where $\sigma$ is the rotation velocity of the interstellar medium
(ISM). From the ISM kinematics,
the rotation velocities of the nearby dIrrs are known to be
typically 30 km s$^{-1}$ (e.g., Mateo 1998). On the other hand,
the cooling timescale for the heated gas is estimated as
\begin{eqnarray}
t_{\rm cool}  \equiv  \frac{3k_{\rm B}T}{2n\Lambda_{\rm cool}},
\end{eqnarray}
where $k_{\rm B}$ and $\Lambda_{\rm cool}$ are the Boltzmann constant and the
cooling function,
respectively. Considering that the metal-line cooling is effective
at $T\sim 10^6$ K,
\begin{eqnarray}
t_{\rm cool}\simeq 7\times 10^8\left(\frac{\zeta}{0.1}\right)^{-1}
\;{\rm yr},\label{cool}
\end{eqnarray}
 where
$\zeta$ is the metallicity normalized by the abundance of the
solar system
(the cooling function at $T=10^6$ K is
$\Lambda_{\rm cool}\simeq 1\times 10^{-22}\zeta$ erg cm$^3$ s$^{-1}$;
Gaetz, Salpeter 1983; see also Raymond et al.\ 1976).
The typical metallicity of a dIrr is $\zeta\sim 0.1$.
Since the cooling timescale is longer than the dynamical timescale,
the time interval of the star-formation activity is determined by
the cooling timescale.
According to the  above estimation,
the cooling time is longer than the crossing time estimated
in the previous subsection.

Here, we should note that the mixing between the hot 
($T\sim 10^6$ K) and
the cool ($T\ltsim 10^4$ K) components makes the
cooling time much shorter. The mixing may produce gas with a
temperature of $\sim 10^5$ K (Begelman, Fabian 1990;
Slavin et al.\ 1993), where the cooling
is an order of magnitude more efficient than that at $10^6$ K.
Thus, the cooling time may be much shorter than that estimated in
equation (\ref{cool}), if the mixing is sufficiently efficient.
According to Roy and Kunth (1995), the timescale of the mixing
between the hot and warm/cold gas is on
the order of  $10^8$ yr, which is longer than the crossing
time. If the mixing is more efficient, the cooling
times may become comparable to, or shorter than, the crossing
time.

Evaporation of the cold gas may can contribute to the cooling
of hot gas. Indeed, equation (\ref{cond})
indicates that evaporation of the hot gas occurs within a short
timescale. Thus, the cooling time may become much shorter. However,
even if a short cooling time is realized, the interval timescale
of the star-formation activity cannot be shorter
than the dynamical time, according to equation (\ref{interval}).
Thus, $t_{\rm int}\gtsim 3\times 10^7$ yr, which is comparable
to, or larger than, the crossing timescale. If $t_{\rm int}$ is
as short as $t_{\rm cross}$, the property of the star-formation
activity may not be intermittent, but continuous. However, the
observed star-formation
activities of a dIrr sample supports the intermittent nature
(subsection 2.4).

The cooling timescale is also important in the context of
the observation of IC 10 by Sait\={o} et al.\ (1992). If the
infalling clouds observed by them originate from the
interstellar gas in IC 10, once heated and flowed out of the
galaxy, the gas must have experienced cooling
(Shapiro, Field 1976; Ikeuchi 1988). The star-formation
activities may be triggered by such clouds. We note that if such
a mechanism
works effectively, an extended dark matter potential is
necessary to prevent the heated gas from escaping.

\subsection{Nature of Star Formation Activity}

In the context of the propagating star-formation model, 
the relation between $t_{\rm cross}$ and
$t_{\rm int}={\rm max}(t_{\rm cool},~t_{\rm dyn})$
determines the nature of the star-formation activity.
If $t_{\rm cross}<t_{\rm int}$, a chain of star-formation
activities affects the whole galactic region and prevents new
star formation activities. In this case, new star-formation
activities emerge after the cooling and dynamical collapse of
the heated gas. Thus, the star-formation activities are
intermittent. On the contrary, when
$t_{\rm cross}>t_{\rm int}$, continuous star-formation
activity is possible, since the
cooling becomes efficient in the crossing time.

Equations (\ref{cross}) and (\ref{cool}) indicate that
$t_{\rm cross}<t_{\rm int}(=t_{\rm cool})$ in dIrrs.
This means that continuous
star formation is not possible, because the stellar feedback
affects the system globally before the heated gas is cooled.
Thus, dIrrs show intermittent star-formation activities and
the timescale of intermittence is
$\sim 7\times 10^8~{\rm  yr}$ according the cooling time. The
Gyr-timescale intermittence roughly agrees with
recent observational results for the star-formation histories
derived from the stellar color-magnitude diagrams
(e.g., Aparicio 1999; Grebel 1999). The existence of the
post-starburst dIrrs also supports the Gyr-timescale
intermittence (Marlowe et al.\ 1995).

The intermittent nature of star formation in a small-sized
galaxy is simulated in Gerola et al.\ (1980) by using a
self-propagating model of star formation. However, we newly
stress that the intermittent nature emerges due to the
interplay between the long cooling timescale and the small
size. Another important point in this paper is the new mechanism
for the short-term variation of star-formation activities.
Recently, Kamaya and Takeuchi (1997) proposed a mechanism for
the short-term variation of star formation activities based
on the limit-cycle model of interstellar medium proposed by
Ikeuchi and Tomita (1983) and Ikeuchi et al.\
(1984).

However, the mixing of interstellar gas should also be
discussed to estimate the cooling time.
We have seen that the cooling timescale can become much
shorter if the mixing is sufficiently efficient (subsection 2.2).
Thus, we only qualitatively conclude here that the {\it small size}
and {\it low metallicity} can show the intermittent star-formation
activity. The efficiency of the mixing should be
investigated in detail in the future. The physical processes
responsible for the mixing are listed along with their timescales
in Roy and Kunth (1995).

The mass loss from dwarf galaxies may affect their evolutions
(e.g, Dekel, Silk 1986; Heckman et al.\ 1990), since the
thermal speed of gas with $T=10^6$ K ($\sim 100$ km s$^{-1}$)
exceeds the escape velocity of
the dwarf systems. However, the efficiency of the mass loss
depends on the size of the dark-matter halo. Though dwarf
galaxies are expected to have extended dark-matter halos by
the analogy with giant galaxies, the size of the halo is
observationally unknown. Thus, it is difficult to constrain the
efficiency of mass loss. Observationally, there is evidence
that some dwarf galaxies in post-starburst phases are still
gas-rich (e.g., Marlowe et al.\ 1995). Thus, the dwarf galaxies
do not necessarily lose all of their gas after their star-formation
activities. A recent theoretical simulation of mass loss from dwarf
galaxies has begun to reveal the mass-loss efficiency (e.g.,
Mac-Low, Ferrara 1999).

For the parameters typical of spiral galaxies ($\zeta\sim 1$ and
$R\sim 10$ kpc), $t_{\rm cross}>t_{\rm cool}$ and 
$t_{\rm cross}>t_{\rm dyn}$ (the typical dynamical timescale of
spiral galaxies is $10^8$ yr). This means 
in the context of the propagating star formation that
the next star formation is possible before propagation
to the whole galaxy. Thus, continuous star-formation activity
in a spiral galaxy is possible in the context of the propagating
star formation.

\subsection{Comparison with Observations}

The intermittent nature of star formation will be observed
as a scatter of the star-formation activity in the dIrr sample.
For giant spiral galaxies, Tomita et al.\ (1996) and 
Kamaya and Takeuchi (1997) regarded the scatter of
the far-infrared-to-optical ratio in the spiral sample as
the time variation of present star-formation activities.
The  H$\alpha$ equivalent widths of spiral galaxies also show
a variety of star-formation activities (Kennicutt et al.\ 1994).
Applying their discussions to dIrrs, we expect that the
dIrr sample shows a variety of star formation activities if
the intermittent nature of star formation activity is realized
in dIrrs.

Indeed, from the {\it UBV} colors of
Marlowe et al.\ (1995)'s dIrr sample show various levels
of star-formation activities. The equivalent widths of
H$\alpha$ also indicate a wide range of present star formation
activities of dIrrs (Marlowe et al.\ 1999).
Thus, the intermittent nature of the star formation is supported
by observations. To confirm the mechanism of the intermittence
proposed in this paper, further examinations (modeling or
observations) are needed. The correlation between the present 
star formation rate at present and that 
averaged over the past 1 Gyr (Brosch et al.\ 1998) may be a
key for the examination.

\section{Summary and Implications}

\subsection{Summary}

In this paper, we have examined the nature of the star-formation
activities of dwarf irregular galaxies (dIrrs) by
considering two processes: stellar feedback and cooling.
The former is the heating process by stars (supernovae,
stellar winds, ultraviolet radiations, thermal conduction, etc.),
and the latter is important to initiate the next star-formation
activity.

First, we phenomenologically applied the observed propagation
velocity of star formation to an estimation of the
propagation timescale. The typical timescale
of the propagation is $10^{7\mbox{--}8}$ yr in dIrrs.
Next, we estimated the cooling time of gas heated by the
feedback mechanism. For dIrrs, the typical timescale for the cooling
is nearly 1 Gyr, which is longer than the propagation timescale
mentioned above. Comparing the two timescales, we finally suggested
that the star-formation activity of dIrrs are
intermittent. The small size (i.e., the short propagation
timescale) and the small metallicity
(i.e., the short cooling timescale) are
both responsible for the intermittence. Efficient
interstellar mixing may prevent intermittence,
because it makes the
cooling time shorter by an order of magnitude (subsection 2.1).
Thus, an examination of the mixing efficiency is
important for a thorough understanding of the star-formation
histories of dIrrs.

The intermittence of the star-formation activities is
observationally supported. The {\it UBV} colors and the H$\alpha$
equivalent widths
indicate that there is a wide range of present star-formation
activities of dIrrs. The intermittent character of the
star-formation activities in dIrrs naturally explains the wide
range of present star-formation activities.

\subsection{Implications}

The above results imply the following:

[1] The gas-consumption timescale is
important for the star-formation histories of dwarf galaxies.
If the gas-consumption timescale is shorter than the
crossing timescale during the initial burst of star formation,
the galaxy only experiences the initial star formation.
The physical properties of the galaxy are determined by the
initial star formation, as discussed in
Hirashita et al.\ (1998).
On the contrary, if the gas-consumption timescale is longer
than the crossing time, intermittent star formation
is possible. In this case, the scenario in
Hirashita et al.\ (1998) is never applicable. The dark-matter
content in a dwarf galaxy is also important
in determining the star-formation histories according to
Hirashita et al.\ (1998; see also Mac Low, Ferrara 1999).

[2] Since the small metallicity is responsible for the
long cooling time, the intermittence becomes more important
as the gas is more primordial. Thus, during the formation epoch of
galaxies, the timescale of metal enrichment is important
for the efficiency of cooling. If only cooling by free-free
radiation is considered to estimate the
metal-free cooling timescale, the cooling time
becomes 4 Gyr (with the cooling rate at $T\sim 10^6$ K of
$\Lambda_{\rm cool}\sim 2\times 10^{-24}$ erg s$^{-1}$ cm$^3$;
Spitzer 1978). This gives an upper bound for the timescale
of the intermittence.
We note that the diverse star-formation histories of dIrrs
 (Schulte-Ladbeck, Hopp 1998)
should be studied in light of the metal-formation history.

[3] The dIrrs in the sample of
Brosch et al.\ (1998b) show no evidence of the
propagating star formation in the morphologies of their
star forming regions. We have shown that even if propagating
star formation really occurs in dwarf galaxies, the duration
of the propagating phase is short compared with the cooling
time. Thus, it may be natural that some dIrrs
show no evidence of propagating star formation.
To give a clear answer as to whether propagating star formation
takes place in dIrrs, more extensive studies are necessary.

\par
\vspace{1pc}\par

We first thank the anonymous referee for careful reading and
making valuable comments, which improved this paper very
much. We offer our gratitude to Dr.\ S. Mineshige for continuous
encouragement. We are grateful to Drs.\ A. Yonehara,
T.~T.~Takeuchi, A. Tomita,
and H.~Kamaya for useful discussions.
This work was supported by the Research Fellowship of the Japan
Society for the Promotion of Science for Young Scientists.
We made extensive use of the NASA's Astrophysics
Data System Abstract Service (ADS).

\section*{References}
\small
\re
Aparicio A.\ 1999, in The Stellar Content of Local Group
Galaxies, IAU Symposium No.\ 192, ed P.\ Whitelock, R.\ Cannon
(ASP, San Francisco) p304
\re
Begelman M.C., Fabian A.C.\ 1990, MNRAS 244, L26
\re
Binney J., Tremaine S.\ 1987, Galactic Dynamics (Princeton
University Press, Princeton) p37
\re
Brosch N., Heller A., Almoznino E.\ 1998a, ApJ 504, 720
\re
Brosch N., Heller A., Almoznino E.\ 1998b, MNRAS 300, 1091
\re
Cowie L.L., McKee C.F.\ 1977, ApJ 211, 135
\re
Cox D.P., Smith B.W.\ 1974, ApJ 189, L105
\re
Dekel A., Silk J.\ 1986, ApJ 303, 39
\re
Dopita M.A., Mathewson D.S., Ford V.L.\
                       1985, ApJ 297, 599
\re
Draine B.T., Giuliani J.L. Jr 1984, ApJ 281, 690
\re
Elmegreen B.G., Elmegreen D.M.\ 1978,
                          ApJ 220, 1051
\re
Feitzinger J.V., Glassgold A.E.,
        Gerola H., Seiden P.E. 1981, A\&A 98, 371
\re
Ferri\`{e}re K.\ 1998, ApJ 503, 700
\re
Gaetz T.J., Salpeter E.E.\ 1983, ApJS 52, 155
\re
Gerola H., Seiden P.E.\ 1978, ApJ 223, 129
\re
Gerola H., Seiden P.E.,
                   Schulman L.S.\ 1980, ApJ 242, 517
\re
Grebel E. K.\ 1999, in The Stellar Content of Local Group
Galaxies, ed P.\ Whitelock, R.\ Cannon (ASP, San Francisco)
p17
\re
Habe A., Ikeuchi S.\ 1980, Prog.\ Theor.\ Phys.\ 64, 1995
\re
Heckman T.M., Armus L., Miley G.K.\ 1990, ApJS 74, 833
\re
Hirashita H.\ 1999, ApJ 520, 607
\re
Hirashita H., Kamaya H., Mineshige S.\ 1997, MNRAS 290, L33
\re
Hirashita H., Takeuchi T.T., Tamura N.\ 1998, ApJ 504, L83
\re
Hodge P.W.\ 1983, AJ 88, 1323
\re
Ikeuchi S.\ 1988, Fundam.\ Cosmic Phys.\ 12, 255
\re
Ikeuchi S., Habe A., Tanaka Y.D.\ 1984, MNRAS 207, 909
\re
Ikeuchi S., Tomita H.\ 1983, PASJ 35, 77
\re
Kamaya H.\ 1998, AJ 116, 1719
\re
Kamaya H., Takeuchi T.T.\ 1997, PASJ 49, 271
\re
Kennicutt R.C. Jr, Tamblyn P., Congdon C.W.\ 1994, ApJ 435, 22
\re
Mac Low M.-M., Ferrara A.\ 1999, ApJ 513, 142
\re
Marlowe A.T., Heckman T.M., Wyse R.F.G., Schommer R.\ 1995,
ApJ 438, 563
\re
Marlowe A.T., Meurer G.R., Heckman T.M.\ 1999, ApJ 522, 183
\re
Mateo M.\ 1998, ARA\&A 36, 435
\re
McKee C.F., Ostriker J.P.\ 1977, ApJ 218, 148
\re
Mueller M.W., Arnett W.D.\ 1976, ApJ
                        210, 670
%\re
%Neukirch T., Feitzinger J.V. 1988, MNRAS 235, 1343
\re
Nozakura T., Ikeuchi S.\ 1988, ApJ 333, 68
\re
Raymond J.C., Cox D.P., Smith B.W. 1976, ApJ 204, 290
\re
Roy J.-R., Kunth D. 1995, A\&A 294, 432
\re
Sait\={o} M., Sasaki M., Ohta K., Yamada T.\
                      1992, PASJ 44, 593
\re
Scalo J., V\'{a}zquez-Semadeni E., Chappell D., Passot T.\
1998, ApJ 504, 835
\re
Schulte-Ladbeck R.E., Hopp U.\ 1998, AJ 116, 2886
\re
Shapiro P.R., Field G.B.\ 1976, ApJ 205, 762
\re
Shostak G.S., Skillman E.D.\ 1989, A\&A 214, 33
\re
Slavin J.D., Shull J.M., Begelman M.C.\ 1993, ApJ 407,
83
\re
Spaans M., Norman C.A. 1997, ApJ 483, 87
\re
Spitzer L. Jr 1956, Physics of Fully Ionized Gases (Interscience,
London) p87
\re
Spitzer L. Jr 1978, Physical Processes in the Interstellar
Medium (Wiley, New York) ch13
\re
Tomisaka K., Bregman J. N.\ 1993, PASJ 45, 513
\newpage
\re
Tomita A., Ohta K., Nakanishi K., Takeuchi T.T., Sait\={o} M.\
1998, AJ 116, 131
\re
Tomita A., Tomita Y., Sait\={o} M.\ 1996, PASJ 48, 285
\re
van den Bergh S.\ 1981, AJ 86, 1464
\re
Weaver R., McCray R., Castor J., Shapiro P., Moore R.\
1977, ApJ 218, 377

%\label{last}

\end{document}